\documentclass[floatfix,twocolumn]{revtex4}
\usepackage{amsmath}
\usepackage{color}
\usepackage{amssymb}

\usepackage{graphicx}

\begin{document}

\title{Low temperature specific heat and thermal conductivity in superconducting \bf UTe$_2$ }

\author{V.P.Mineev$^{1,2}$}
\affiliation{$^1$Universite Grenoble Alpes, CEA, IRIG, PHELIQS, F-38000 Grenoble, France\\
$^2$Landau Institute for Theoretical Physics, 142432 Chernogolovka, Russia}

\begin{abstract}
The  measurements  (Phys.Rev.B {\bf 100}, 220504(R) (2019)) do not detect noticeable  thermal 
conductivity in superconducting  UTe$_2$  in the T=0 limit. At the same time the same crystals  exhibit a large residual density of states  comparable with its normal state value.
The improvement of samples quality  leads to the augmentation of critical temperature of transition to the superconducting state accompanied by the decreasing of residual specific heat ratio $C(T)/T$ at $T\to 0$. 
There is presented an analytic derivation of this inverse correlation property in concrete cases of 
triplet superconducting states with node-less  and point-nodes order parameters allowed by symmetry in UTe$_2$.
The obtained explicit formulas for the residual thermal conductivity are in reasonable correspondence with observations.
\end{abstract}

\date{\today}
\maketitle

\section{Introduction}

The recently discovered uranium superconductor UTe$_2$ \cite{Ran-Science,Aoki2019} has rather unusual qualities such as  reentrant superconductivity at extremely high magnetic fields \cite{Ran-NatPhys,Knebel2019}.  
In the early experiments  the bulk superconductivity  appeared at $T_c\approx 1.6$ K and characterised
 by the sharp specific heat jump larger than the weak coupling BCS value $\Delta C/\gamma T_c=1.43$ and by
remarkably large residual ratio $C/T=\gamma_{sc}$   at $T\to 0$  which is equivalent to  approximately one half of its normal state $\gamma_n$  value\cite{Ran-Science,Aoki2019}.The little variation of residual density of states in UTe$_2$ between the samples with slightly different $T_c$ led to the assumption that
the large residual electronic density of states is likely an intrinsic, disorder-insensitive property of UTe$_2$ as if its superconducting state was 
nonunitary such that only half of the electrons 
participate in spin-triplet pairing, while the remainder continue to form a normal Fermi liquid. 
At the same time, 
 the thermal conductivity measurements \cite{Paglione2019} do not show noticeable  thermal 
conductivity in the T=0 limit for the same crystals that exhibit a large residual density of states. 
Then, however, the higher  $T_c\approx 1.77 $ K was registered and there was established the inverse correlation between $T_c$ and 
$\gamma_{sc}$ \cite{Cairns2020,Aoki2020}. At last in the highest quality crystals with RRR=57 there was found the superconducting transition temperature $T_c=2 K$ and the lowest  residual heat capacity with $\gamma_{sc}$ noticeably smaller than $\gamma_n$ \cite{Rosa2021,Aoki2021}.

The inverse  correlation between $T_c$ and $\gamma_{sc}$ has also been observed in other unconventional superconductors  UPt$_3$ \cite{Fisher1989} and CeCoIn$_5$ \cite{Tanatar2005}. The residual density of states originates from the electron  scattering on impurities and crystal imperfections forming  so called gapless superconducting state. Discovered first by Abrikosov and Gor'kov \cite{AbrGor1960} in conventional superconductors with paramagnetic impurities
this phenomenon in unconventional superconductors with ordinary impurities  has been investigated in the papers \cite{Hirschfeld1988, Barash1997} ( see also the textbook \cite{Mineev1999}). 
The temperature dependence of the thermal conductivity in unconventional superconductors including  T=0 limit has been studied in articles\cite{Arfi1988,Hirschfeld1988,Graf1996} expouned in the textbook \cite{Mineev1999}.

The superconducting properties of UTe$_2$ are strongly dependent on the synthesis route. The relatively small RRR ratio   points on the significant presence of some crystal imperfections  probably originating from the non-stoichiometric composition of this material. These type of imperfections not spoiling the crystal periodicity on the large scales serve as the  electron scattering centres.
Here, I consider  the property of inverse correlation between the critical temperature and residual density of states  as well the zero temperature limit of the thermal conductivity based on the theory of resonant scattering  in concrete application to the possible superconducting states in UTe$_2$.   
The problem of compatibility of finite residual density of states and small residual thermal conductivity is discussed.

The paper is organised as follows. In the next section
there is obtained the explicit expressions for the residual density of states in 
triplet superconducting states with node-less  and point-nodes order parameters.
It is shown that in both cases  the residual density of states  in practically pure crystal may represent a significant part of 
its normal state value. 
Then  in the next section there is established the T=0 limit  of thermal conductivity. 
The residual thermal conductivity  in clean enough specimens in both cases of superconducting states with node-less and point-nodes order parameter turns out to be comparable with with values observed experimentally \cite{Paglione2019}.

\section{Residual density of states}

UTe$_2$ has an orthorhombic structure.  In  view of abnormally large upper critical field in UTe$_2$ one must discuss  the superconducting states with triplet pairing. 
The  superconducting states  with triplet pairing  in a metal with orthorhombic structure are related to one of four different representations 
of the point group $D_{2h}$:  $A_{u}, B_{1u},B_{2u}, B_{3u}$. The corresponding order parameters of superconducting states are
\begin{eqnarray}
{\bf d}_{0}({\bf k})=\Delta_0(\varphi_{0x}\hat x+\varphi_{0y} \hat y+\varphi_{0z} \hat z),\nonumber\\
{\bf d}_{B1}({\bf k})=\Delta_1(\varphi_{1y}\hat x+\varphi_{1x} \hat y+\varphi_{1w} \hat z),\nonumber\\
{\bf d}_{B2}({\bf k})=\Delta_2(\varphi_{2z}\hat x+\varphi_{2w} \hat y+\varphi_{2x} \hat z),\nonumber\\
{\bf d}_{B3}({\bf k})=\Delta_3(\varphi_{3w} \hat x+\varphi_{3z} \hat y+\varphi_{3y} \hat z).\nonumber
\end{eqnarray}
Here,  $\Delta_i$, $i=0,1,2,3$ are the  order parameter amplitudes. The functions $\varphi_{ix}({\bf k}),\varphi_{iy}({\bf k}),\varphi_{iz}({\bf k}),\varphi_{iw}({\bf k})$ transform as $k_x,k_y,k_z,k_xk_yk_z$ correspondingly. The order parameter of $A$ state has finite value at any point on the Fermi surface, whereas the order parameter of $B_i$ state for $i = 1,2,3$ has the point nodes 
at $\hat{\bf k}$   along $ z,y,x$ direction  correspondingly. To do  the calculations analytically we will consider the superconducting states with more simple order parameters possessing the same properties:
 \begin{equation}
{\bf d}_{A}({\bf k})=\Delta_A(\hat k_x\hat x+\hat k_y\hat y+\hat k_z\hat z),
\label{A}
\end{equation}
for unit representation $A$
and
\begin{equation}
{\bf d}_{B}({\bf k})=\Delta_{B}(\hat k_y\hat x+\hat k_x\hat y).~~~~~~
\label{B}
\end{equation}
for  non-unit representations $B_i$.

The critical temperature in a dirty unconventional superconductor is  determined by the Abrikosov-Gorkov equation
 \begin{equation}
 \ln\frac{T_{c0}}{T_c}=\psi\left ( \frac{\Gamma}{2\pi T_c}+ \frac{1}{2}\right)-
 \psi\left ( \frac{1}{2} \right),
 \end{equation}
 where $\Gamma$ is the scattering rate of electrons on impurities.
The superconductivity disappears at
\begin{equation}
\Gamma_c=\frac{\pi}{2\gamma}T_{co}=\frac{\Delta_{00}}{2}, 
\end{equation}
where $T_{c0}$ is the critical temperature in the absence of impurities, $\Delta_{00}$ is the order parameter amplitude in the absence of impurities at T=0, $\gamma=e^C\approx 1.78$, $C\approx 0.577$  is the Euler constant.

Found in the Born approximation \cite{AbrGor1960} in the limit $T\to 0$ the finite density of states at Fermi level  appears in very dirty limit starting from the scattering rate 
\begin{equation}
\Gamma_g={2}\Gamma_c\exp\left (-\frac{\pi}{4}  \right )\approx 0.91\Gamma_c.
\label{gamma}
\end{equation}
We will be interested in the opposite situation of almost clean superconductor, where the gapless superconducting state is realised in case of resonant scattering on impurities with the scattering rate
 \begin{equation}
 \Gamma=\frac{n_i}{\pi N_0}\ll \Gamma_c.
 \end{equation}
Here, $N_0$ is the electron density of states at the Fermi level on one spin projection. 

For small $\Gamma$  the critical temperature is
\begin{equation}
 T_c \cong T_{c0}-\frac{\pi}{4}\Gamma,
 \end{equation}
 that  is 
 \begin{equation}
\Gamma=\frac{8\gamma}{\pi^2}\Gamma_c\tau, 
\label{Gaa}
\end{equation}
where
 $$
 \tau=\frac {T_{c0}-T_c}{T_{c0}}\ll 1.
 $$
Let us write now the residual density of states for the superconducting states corresponding to unit representation (\ref{A}) and to anyone of non-unit representation (\ref{B}).

\subsection{ State A}

The nonzero density of states at $E=0$ is derived following procedure described in \cite{Mineev2001}
\begin{equation}
N_0(E=0)=N_0\left [
\frac{\left (\frac{\Gamma}{\Delta_0}\right )^2+\frac{\Gamma}{\Delta_0}\sqrt{\left (\frac{\Gamma}{\Delta_0}\right )^2+4}}
{2+\left (\frac{\Gamma}{\Delta_0}\right )^2+\frac{\Gamma}{\Delta_0}\sqrt{\left (\frac{\Gamma}{\Delta_0}\right )^2+4}}
\right ]^{1/2}.
\label{N00}
\end{equation}
The order parameter amplitude at $T\to 0$ and small $\Gamma$   is \cite{Mineev2001} 
\begin{equation}
{\Delta_0}\approx{\Delta_{00}}\left (1- \frac{\pi}{4}\frac{\Gamma}{\Gamma_c}\right ).
\label{OP}
\end{equation}
Hence, at small $\Gamma$ the residual density of states is
  \begin{equation}
  N_0(E=0) \approx\sqrt{\frac{\Gamma}{\Delta_{00}}}=\frac{2\sqrt{ \gamma}}{\pi}N_0\sqrt{\tau}.
  \end{equation}

\subsection{State B}

The calculation of the residual density of states for a state $B_i$ coincides with corresponding treatment for the superconducting state with structure of $A$-phase of superfluid $^3$He presented in the textbook \cite{Mineev1999}. It is
\begin{equation}
 N_0(E=0)=N_0\sqrt{\frac{\pi\Gamma}{2\Delta_{0}}}.
\end{equation}
 In case of weak disorder the difference between $\Delta_{00}$ and $\Delta_0$ is small.  Hence,
 \begin{equation}
 N_0(E=0)\approx N_0\sqrt{\frac{\pi\Gamma}{2\Delta_{00}}}=\sqrt{\frac{2\gamma}{\pi}}N_0\sqrt{\tau}. 
\end{equation}

  We see that the zero temperature density of states can be finite even at small enough $\tau$. Moreover, in the samples with higher critical temperature  
   the residual density of states $N_0(E=0)$ is smaller that is in qualitative correspondence with experimental observations \cite{Rosa2021,Aoki2021}.

 \section{Low temperature thermal conductivity }

 In neglect of vertex corrections the thermal conductivity in a superconducting state in $i=x,y,z$-direction is \cite{Hirschfeld1988,Graf1996,Mineev1999}
 \begin{eqnarray}
 \kappa_{i}=\frac{N_0v^2_F}{4T^2}\int_0^\infty\frac{dE~E^2}{\cosh^2(E/2T)}~~~~~~~~~~~\nonumber\\
\times\frac{1}{{\text Re}~t{\text Im}~t}\int\frac{d\Omega}{4\pi}{\hat k_i^2}
 {\text Re}\frac{|t|^2+t^2-2|{\bf d}({\bf k})|^2}{\sqrt{t^2-|{\bf d}({\bf k})|^2}}.
 \label{kappa}
 \end{eqnarray}

 \subsection{ State A} 
 
Here $t=t(E)$ is determined by the equation
 \begin{equation}
 t=E-\frac{\Gamma}{t}\sqrt{\Delta_0^2-t^2}.
 \end{equation}
 The solution of this equation at $\Gamma\ll\Delta_0$ and $E\to 0$ is 
  \begin{equation}
 t\approx \frac{E}{2}+i\sqrt{\Delta_0\Gamma}.
 \end{equation}
  The integral over angles in Eq.(\ref{kappa})   at $E \ll \sqrt{\Delta_0\Gamma}$ is
 \begin{eqnarray}
 \frac{1}{{\text Re}~t{\text Im}~t}{\text Re} \int\frac{d\Omega}{4\pi}{\hat k_i^2}
\frac{|t|^2+t^2-2|{\bf d}({\bf k})|^2}{\sqrt{t^2-|{\bf d}({\bf k})|^2}}\nonumber\\
\approx 2\int\frac{d\Omega}{4\pi}\hat k_i^2\frac{\Delta_0\Gamma}{(\Delta_0^2+\Delta_0\Gamma)^{3/2}}
\approx \frac{2\Gamma}{3\Delta_0^2}.
 \end{eqnarray}

 Then the thermal conductivity near the zero temperature is
 \begin{equation}
 \kappa=\frac{N_0v_F^2\Gamma}{6T^2\Delta^2_0}\int_o^\infty\frac{E^2dE}{\cosh^2\frac{E}{2t}}=\frac{N_0v_F^2\Gamma T}{6\Delta^2_0}I,
 \label{1}
 \end{equation}
 where
 $$
 I=\int_0^\infty\frac{z^2dz}{\cosh^2\frac{z}{2}}=\frac{2\pi^2}{3}.
 $$
 
  Comparing the obtained result with
  the thermal conductivity in normal state
 \begin{equation}
 \kappa_n=
 \frac {N_0v_F^2T}{6 \Gamma}I,
 \end{equation}
 we find
 \begin{equation}
 \frac{\kappa_A(T\to 0)}{\kappa_n(T=T_c)}=\left (\frac{\Gamma}{\Delta_0}\right )^2\frac{T}{T_c}.
 \label{ratio}
 \end{equation}

   \subsection{State B}

The equation for the function $t(E)$ in the state $B$ is
\begin{equation}
t=E+2i\Gamma\Delta_0\left (t\ln\frac{t+\Delta_0}{t-\Delta_0}\right )^{-1}.
\end{equation}
 The solution of this equation at $\Gamma\ll\Delta$ and $E\to 0$ is
 \begin{equation}
  t\approx \frac{E}{2}+i\sqrt{\frac{2}{\pi}\Delta_0\Gamma}
 \end{equation}
 The integral over angles in Eq.(\ref{kappa})   at $E \ll \sqrt{\Delta_0\Gamma}$ is
 \begin{eqnarray}
 \frac{1}{{\text Re}~t{\text Im}~t}{\text Re} \int\frac{d\Omega}{4\pi}{\hat k_i^2}
\frac{|t|^2+t^2-2|{\bf d}({\bf k})|^2}{\sqrt{t^2-|{\bf d}({\bf k})|^2}}\nonumber\\
\approx 
2\int\frac{d\Omega}{4\pi}\hat k_i^2\frac{\frac{2\Delta_0\Gamma}{\pi}}{(\Delta_0^2(\hat k_x^2+\hat k_y^2)+\frac{2\Delta_0\Gamma}{\pi})^{3/2}}\nonumber\\
\approx \left \{\begin{array}{l}     \frac{2\sqrt{2}}{{\sqrt\pi}}\frac{\sqrt{\Gamma}}{\Delta_0^{3/2}} , ~~~i=z,  ~~~~~~~~ \\
~~~~\\
\frac{\Gamma}{\Delta_0^2},~~~~~~~~~i=x,y. \end{array} \right.~~~~~~~~
 \end{eqnarray}
 Hence, the residual thermal conductivity in the direction parallel to the point node direction is
 \begin{equation}
 \kappa_{B\parallel}=\frac{N_0v_F^2T}{\sqrt{2\pi}}
 \frac{\sqrt{\Gamma}}{\Delta_0^{3/2}}I,
 \label{2}
 \end{equation}
and in the perpendicular direction
 \begin{equation}
 \kappa_{B\perp}
 =\frac{N_0v_F^2T}{4}\frac{\Gamma}{\Delta_0^2}I.
 \label{3}
 \end{equation}
 The ratios of residual thermal conductivity in parallel and perpendicular directions to the normal state  thermal conductivity at $T=T_c$ are
\begin{equation}
 \frac{\kappa_{B\parallel(}T\to 0)}{\kappa_n(T=T_c)}=\frac{6}{\sqrt{2\pi}}\left (\frac{\Gamma}{\Delta_0}\right )^{3/2}\frac{T}{T_c},
 \label{parallel}
 \end{equation}
\begin{equation}
 \frac{\kappa_{B\perp}(T\to 0)}{\kappa_n(T=T_c)}=\frac{3}{2}\left (\frac{\Gamma}{\Delta_0}\right )^2\frac{T}{T_c}.
 \label{perp}
 \end{equation}
One can compare these expressions with corresponding ratio  reported in the paper \cite{Paglione2019} at lowest temperature $T=0.05K$ for the specimen S1
\begin{equation}
\left . \frac{\kappa_s(T)}{\kappa_n(T=T_c)}\right |_{exp}\approx 6\times 10^{-2}\frac{T}{T_c}.
\end{equation}
The similar estimation for the specimen S2 at the lowest temperature $T=0.1K$ is
\begin{equation}
\left . \frac{\kappa_s(T)}{\kappa_n(T=T_c)}\right |_{exp}\approx 7\times 10^{-2}\frac{T}{T_c}.
\end{equation}
 The comparison these values with   the theoretical results given by Eqs.(\ref{ratio}), (\ref{parallel}, (\ref{perp}) yields
 \begin{equation}
 \frac {\Gamma}{\Delta_0}\sim 10^{-1}.
 \label{res}
 \end{equation}
 Thus, the residual thermal conductivity  in clean enough specimens in both cases of superconducting states with node-less and point-nodes order parameter turns out to be comparable with with values observed experimentally \cite{Paglione2019}.

\section{Conclusion}
 
 The problems of residual density of states and residual thermal conductivity 
  in triplet superconducting states with node-less and point-nodes order parameters allowed by symmetry in UTe$_2$ have been investigated. There was  shown that in all cases
 the  resonant  scattering on impurities leads to the formation of the gapless superconducting states with large enough density of states at zero energy. 
The increasing  of scattering rate on impurities  suppresses the critical temperature and leads  to the augmentation of the residual density of states. 

The calculations were performed  in clean enough limit when critical temperature of transition to the superconducting states is slightly deviates from the corresponding critical temperature in  perfect crystal, that corresponds to the scattering rate much smaller than the zero temperature amplitude of the order parameter $\Gamma\ll \Delta_0$.
The comparison of the derived explicit formulas for the residual thermal conductivity with corresponding experimental values is in the reasonable correspondence (\ref{res}) with this demand. 
The residual thermal conductivity  in clean enough specimens in both cases of superconducting states with node-less and point-nodes order parameter turns out to be comparable with with values observed experimentally \cite{Paglione2019}.
Still, the superconducting state with point node order parameter is preferable  in view of non-exponential
decreasing of thermal conductivity and the penetration depth at finite temperature  reported in the paper \cite{Paglione2019}.

\acknowledgments

I am indebted to J.Pagleone for the stimulating comment and to D.Aoki for the  attraction of my attention to the recent experimental results.

\end{document}